\documentclass[english,aps,prper,longbibliography,reprint,superscriptaddress]{revtex4-2}   
\usepackage{lipsum}
\usepackage[T1]{fontenc}	% should generally be included for better accented-word behavior
\usepackage[latin9]{inputenc}	% should generally be included for better accent behavior
\usepackage{geometry}		% for controlling page margins
\usepackage{graphicx}
\usepackage{color}
\usepackage[above,below]{placeins}	% allows use of \FloatBarrier command to force section barriers
\usepackage{times}
\usepackage[colorlinks=true,allcolors=blue]{hyperref}
\usepackage{multirow}
\usepackage{rotate}
\usepackage{lineno,xcolor}
\usepackage[colorlinks=true,allcolors=blue]{hyperref}
\usepackage{multirow}
\usepackage{rotate}

\usepackage{amsmath,amssymb}
\usepackage{array}
\usepackage[caption = false]{subfig}

\begin{document}
\title{The Physics Inventory of Quantitative Literacy: \\A tool for assessing mathematical reasoning in introductory physics}

\author{Suzanne White Brahmia}
\affiliation{Department of Physics, University of Washington, Box 351560, Seattle, WA 98195-1560, USA}

\author{Alexis Olsho}
\affiliation{Department of Physics, University of Washington, Box 351560, Seattle, WA 98195-1560, USA}

\author{Trevor I.\ Smith}
\affiliation{Department of Physics \& Astronomy and Department of STEAM Education, Rowan University, 201 Mullica Hill Rd., Glassboro, NJ 08028, USA}%\affiliation{Department of STEAM Education, Rowan University, 201 Mullica Hill Rd., Glassboro, NJ 08028, USA}

\author{Andrew Boudreaux}
\affiliation{Department of Physics \& Astronomy, Western Washington University, 516 High St., Bellingham, WA 98225, USA}

\author{Philip Eaton}
\affiliation{School of Natural Sciences and Mathematics, Stockton University, Galloway, NJ 08205, USA}

\author{Charlotte Zimmerman}
\affiliation{Department of Physics, University of Washington, Box 351560, Seattle, WA 98195-1560, USA}

\begin{abstract}
    One desired outcome of introductory physics instruction is that students will develop facility with reasoning quantitatively about physical phenomena. Little research has been done regarding how students develop the algebraic concepts and skills involved in reasoning productively about physics quantities, which is different from either understanding of physics concepts or problem-solving abilities. We introduce the Physics Inventory of Quantitative Literacy (PIQL) as a tool for measuring \textit{quantitative literacy}, a foundation of mathematical reasoning, in the context of introductory physics. We present the development of the PIQL and evidence of its validity for use in calculus-based introductory physics courses. Unlike \emph{concept} inventories, the PIQL is a \emph{reasoning} inventory, and can be used to assess reasoning over the span of students' instruction in introductory physics. Although mathematical reasoning associated with the PIQL is taught in prior mathematics courses, pre/post test scores reveal that this reasoning isn't readily used by most students in physics, nor does it develop as part of physics instruction---even in courses that use high-quality, research-based curricular materials. As has been the case with many inventories in physics education, we expect use of the PIQL to support the development of instructional strategies and materials---in this case, designed to meet the course objective that all students become quantitatively literate in introductory physics. 
\end{abstract}

\maketitle

\section{Introduction}
\label{sec:intro}

Introductory physics uses familiar mathematics in distinct ways to describe the world and make meaning. To an expert, a physics equation ``tells the story'' of an interaction or process. For example, when reading the equation 
\begin{equation*}
   +10\textrm{ N}=-(20\textrm{ N/m})x,
\end{equation*}
an expert, noting the units and the signs, may consider it to be an example of Hooke's Law. They may then construct a mental story about the force exerted by a spring on a hand as the spring is stretched. The coordinate system is set by the (positive) sign in front of the spring force term which, in the case of a stretched spring, points toward the spring. Although the variable $x$ typically represents the \textit{position} of an object in mechanics, experts recognize that it is being used here to represent the \textit{displacement} of the end of the spring from its equilibrium position. The negative sign accounts for the fact that both the force and the displacement are one-dimensional vector quantities, and they are always in opposite directions for a spring. Putting these insights together, the expert might construct the story of a spring with a stiffness of 20 N/m being stretched by a hand in the negative direction and exerting a force on the hand of 10 N in the positive direction. Part of the challenge of learning physics is developing the ability to decode symbolic representations in this manner. For physicists, equations tell stories, but students don't always recognize these stories even when they ``know'' the prerequisite mathematics. 

Instruction in college physics courses leans heavily on \textit{quantitative literacy}, the interconnected skills, attitudes, and habits of mind that together support the sophisticated use of familiar mathematics for sense-making \cite{thompson2010,ojose2011}. Given the mathematical nature of even introductory-level courses in physics, \textit{Physics Quantitative Literacy (PQL)}, i.e., quantitative literacy in the context of physics, has the potential to be an important learning outcome for all students taking introductory physics. PQL is characterized by the blending of conceptual and procedural mathematics to generate and apply quantitative models of physical phenomena, a skill valued in all STEM fields.

While PQL is a desired outcome of physics instruction, there is a dearth of validated measures of reasoning about quantities and their relationships in physics contexts. Kalinowski and Willoughby developed a \textit{reasoning} inventory to assess college students' scientific reasoning \cite{kalinowski2019}. While their instrument includes several basic proportional reasoning and probability items, its primary focus is on verbal interpretations involving control of variables, hypothesis testing, and correlational reasoning. We have developed the Physics Inventory of Quantitative Literacy (PIQL) to broaden options for assessing quantitative reasoning in physics. The PIQL assesses quantification typically used in introductory physics, with a focus on the mathematical relationships between physics quantities. We intend use of the PIQL in introductory courses to raise awareness among instructors about the extent to which instructional goals surrounding PQL are being met, perhaps in a manner similar to how early concept inventories such as the Force Concept Inventory (FCI) \cite{Hestenes1992} raised awareness that basic concepts were not being understood at the level many instructors assumed. An increased awareness that broad instructional goals are not being met can drive curriculum development to build deeper understanding \cite{Hake1998,VonKorff2016,Madsen2017}. To help facilitate administration, the PIQL is available online for widespread use.

After multiple iterations, the PIQL is in a steady state and we have constructed an argument for its validity in the context of calculus-based introductory physics courses at two selective research universities in the Pacific Northwest, referred to hereafter as \textit{Institution 1} and \textit{Institution 2}. In this paper, we provide evidence of the PIQL's validity as a measure of development of PQL in this population of students \footnote{We recognize that no instrument is ever ``valid'' or ``validated,'' as validity is a quality of the claim being made, not the instrument itself; nor can any claim be \emph{proved} valid. We have attempted to use language that reflects this.}. 
%We recognize that this evidence does not apply to its validity with all other populations, and that physics education research more generally has not always been based on representative samples of students \cite{kanim2020}. In future work, we plan to establish evidence for the validity of the PIQL more broadly, modifying the instrument as necessary based on data collected with more diverse groups of introductory physics students. 
We recognize that this evidence is not representative of all introductory physics students and does not apply to the validity of the PIQL with all other populations. We acknowledge that physics education research more generally has not always been based on representative samples of students \cite{kanim2020}---but like much of this previous work, the analysis and results are informative and useful nonetheless. In future work, we plan to establish evidence for the validity of the PIQL more broadly, modifying the instrument as necessary based on data collected with more diverse groups of introductory physics students.

% Reviewer's Comment: I would restructure the part about the representativeness of the samples to note that you recognize your data are not representative, but like that of much previous work in this field, the analysis still provide useful results to be shared with the field as a whole and goals for future work to better understand the effectiveness of the instrument when deployed with different populations of students.

In the next section, we present the theoretical underpinnings of the PIQL. Section \ref{sec:development} describes the methods used to develop individual items and collect them into a full instrument. The results of both qualitative and quantitative analyses to provide evidence for the validity of the PIQL for use in a calculus-based introductory physics sequence are presented in Section \ref{sec:validation}. We conclude with a discussion of plans to investigate the validity of the PIQL for use in all introductory physics courses and across multiple instructional settings.

\section{Theoretical Foundations}
\label{sec:foundations}

Physics quantitative literacy involves \textit{proceptual} understanding of both the mathematics involved in creating physics quantities and in the quantitative models that relate the quantities to each other. In this section, we describe the role proceptual reasoning plays in the PIQL, how we use the framework of \emph{conceptual blending theory} to interpret proceptual reasoning as a blend of procedural mathematics and physical reasoning, and discuss Sherin's symbolic forms as a poignant framing of proceptual reasoning in introductory physics \cite{Sherin2001}.

Mathematics education researchers Gray and Tall define proceptual understanding as a combination of \textit{pro}cedural mastery and con\textit{ceptual} understanding \cite{gray1994}. They explain that in the context of fractions, for example, ``the symbol $\frac{3}{4}$ stands for both the process of division and the concept of fraction''; that is, a student with a proceptual understanding of fractions would move fluidly between the procedure of dividing 3 by 4, and the physical instantiation of the fraction $\frac{3}{4}$ as a precise quantification of portion. Similarly, a physics student with a proceptual understanding of the quantity torque would move fluidly between the procedure, finding the vector cross product of a position relative to an origin and a force, and conceptualizing the vector product $ \vec{r} \times \vec{F} $ as a quantity unto itself (i.e., as simply $\vec{\tau}$), with its own, important emergent properties and consequences. 

% We frame PIQL as a probe of proceptual algebraic reasoning with physics quantities that is a hallmark of mastery in introductory physics. 
The PIQL assesses proceptual reasoning involving algebraic relationships between physics quantities. Such reasoning characterizes mastery of the content typically presented in introductory, calculus-based physics courses. The PIQL is designed to span a knowledge space \cite{falmagne1990} based on elements of mathematical reasoning ubiquitous in introductory physics. As illustrated in the example in the introduction to this paper, much of the mathematics used in introductory physics courses holds physical significance beyond its strict mathematical meaning. Many (or most) of the PIQL items require proceptual reasoning of the type described above. We regard such reasoning with familiar mathematics---in which ``doing math'' is not separate from ``doing physics''---as meaningful quantitative understanding.  

Conceptual blending theory (CBT) \cite{Fauconnier2002} provides a framework for understanding the integration of mathematical and physical reasoning involved in PQL \cite{Bing2007, Hu2013, VanDenEynde2020}. In their theory, Fauconnier and Turner describe a cognitive process in which a unique mental space is formed from two (or more) separate mental spaces. This blended space can be thought of as a product of the input spaces, rather than a separable sum. According to CBT, development of expert mathematization in physics would occur not through a simple addition of new elements (physics quantities) to an existing cognitive structure (arithmetic or algebra), but rather through the creation of new and independent cognitive spaces. These spaces, in which creative, quantitative analysis of physical phenomena can occur, involves a continuous interdependence of thinking about the mathematical and physical worlds. In the context of the PIQL, we consider proceptual reasoning about physical quantities and the relationships between them to be represented by the blended space, which is an inseparable blend of mathematics reasoning and physical meaning-making.

As we discuss more fully below, the process of inventory development described by Adams and Wieman \cite{adams2010} guided the development of the PIQL. In this process, phase 1 of development involves delineating what the inventory is intended to measure, i.e., establishing the \textit{test construct} \cite{adams2010}. We developed the test construct for the PIQL based on Sherin's theory of symbolic forms, which in turn was developed to explain how successful introductory physics students understand and construct equations \cite{Sherin2001}. Sherin's symbolic forms provides a framework for characterizing the reasoning targeted in PIQL.
\begin{quote}
\dots successful (physics) students learn to understand what equations say in a fundamental sense; they have a feel for expressions, and this understanding guides their work\dots We do students a disservice by treating (physics) conceptual understanding as separate from the use of mathematical notations \cite{Sherin2001}.
\end{quote}

\begin{figure}
  %  \centering
    \includegraphics[width = \columnwidth]{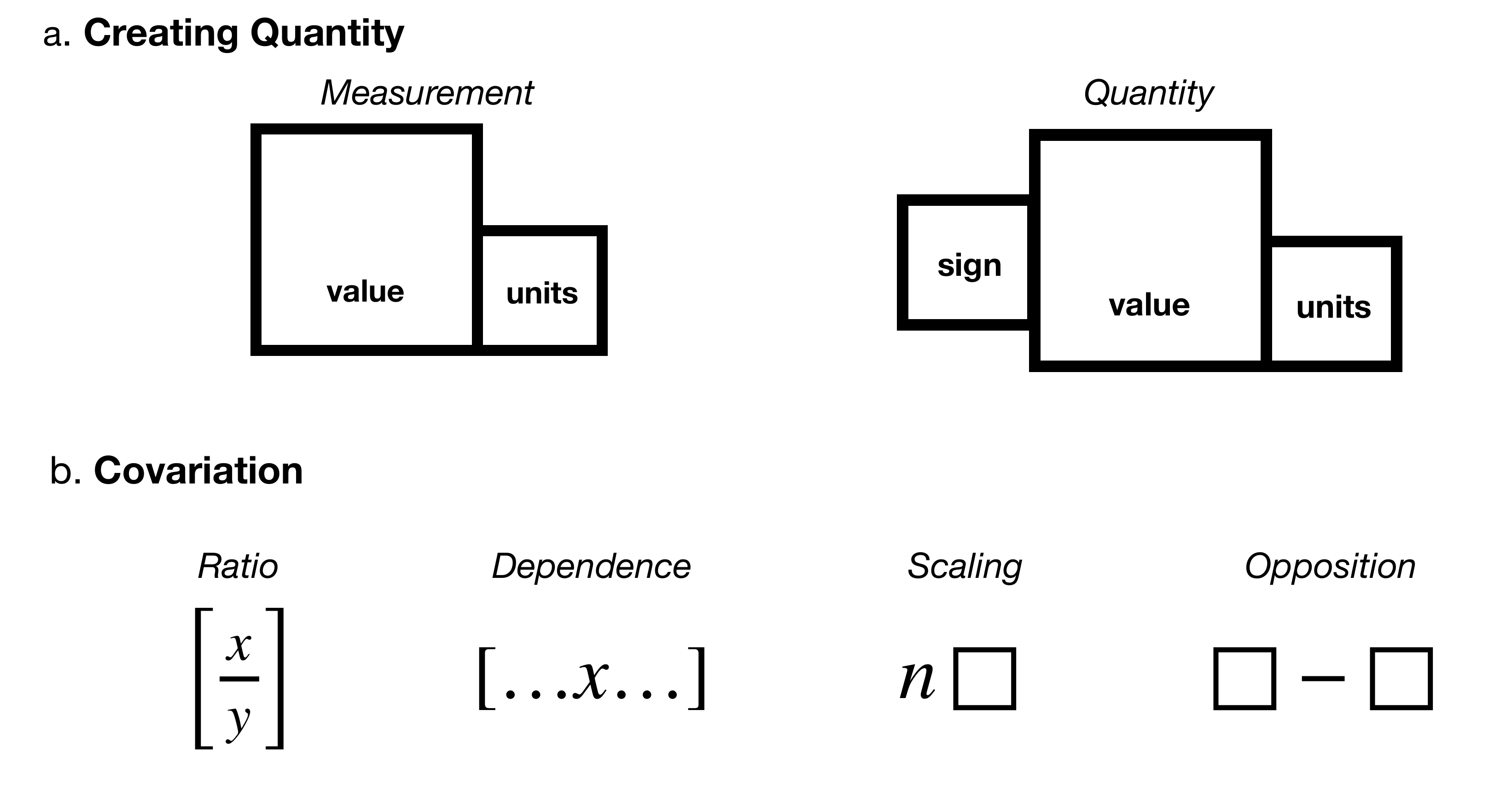}
    \caption{Examples of symbolic forms represented in the PIQL. a) Creating Quantity: Measurement \cite{Dorko2015CalculusUnits} and Quantity \cite{brahmia2019quantification}, and b) Covariation: Ratio, Dependence, Scaling, and Opposition \cite{Sherin2001} .
    }
    \label{fig:symbolic}
\end{figure}

The symbolic forms framework hypothesizes that successful students can develop expert-like conceptual schema with which they associate certain symbol patterns in equations. Sherin developed a list of these symbolic forms, noting that it is not comprehensive, with the intention that subsequent research could help build up a library \cite{Sherin2001}. Fig. \ref{fig:symbolic} shows examples of symbolic forms that are represented in the PIQL, both from Sherin's original work and from more recent work by other researchers \cite{Dorko2015CalculusUnits,brahmia2019quantification}. These forms can be considered a representation of proceptual reasoning in physics; they are the outcome of both a procedural view of related symbols and the physics contexts in which they are relevant. Note that expert reasoning described in the spring example in the introduction to this paper, and in the PIQL item in Fig.~\ref{fig:workProb} rely on the \textit{quantity} and the \textit{scaling} symbolic forms.

Most introductory physics students are less sophisticated in their use of algebraic structures \cite{tuminaro2004}, and come from less privileged backgrounds \cite{kanim2020} than the students in Sherin's study, who were in the last semester of calculus-based physics at an elite institution. As such, we consider symbolic forms to be a \textit{learning objective} of an introductory physics course, rather than characteristic of how typical physics students think. PIQL is designed as a probe to help researchers and instructors meet this objective.

\section{Development of the PIQL}
\label{sec:development}

\begin{figure*}
    \centering
    \includegraphics[width = 0.7\textwidth]{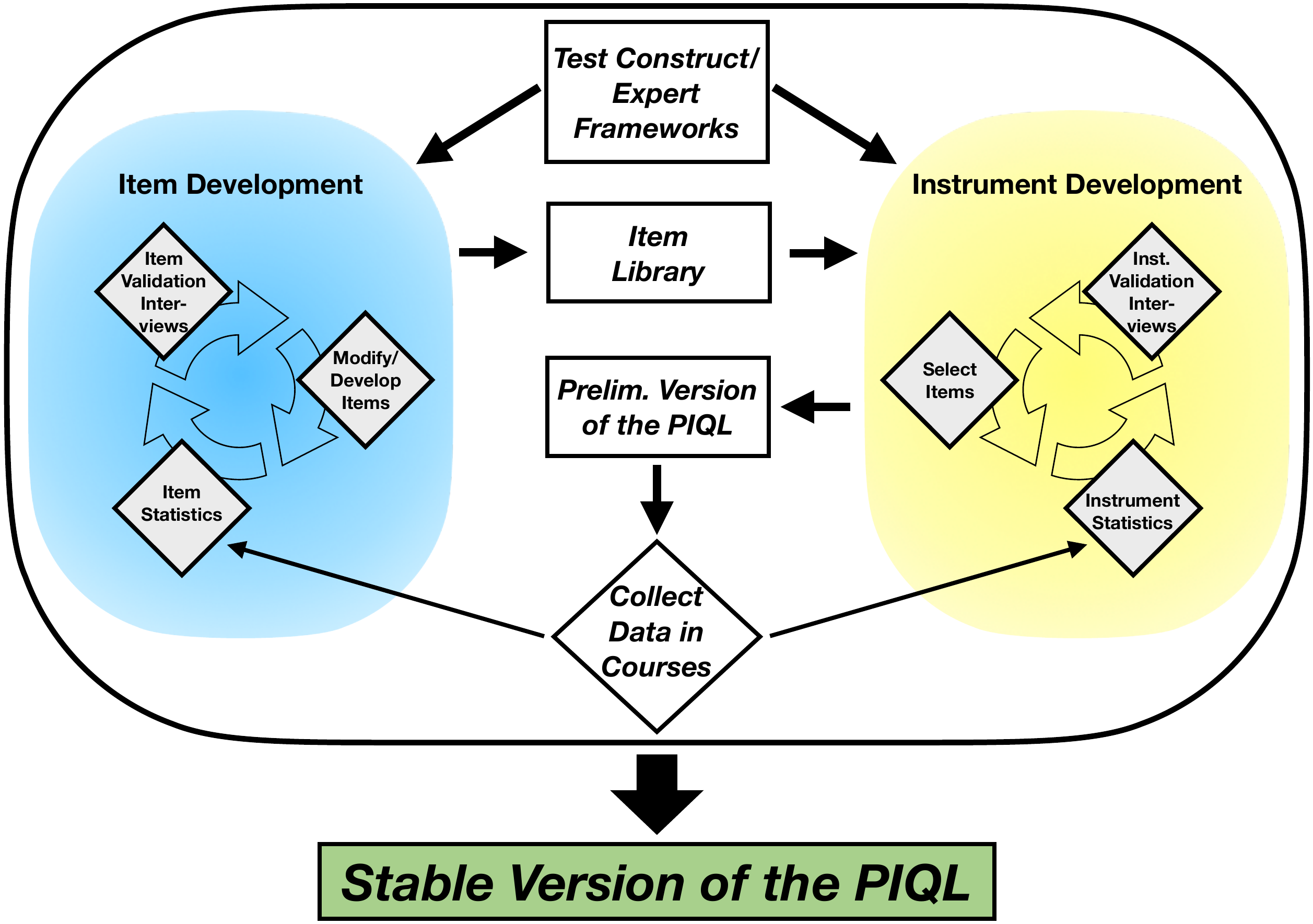}
    \caption{Workflow for developing and revising items, piloting individual items using interviews, assembling versions of the PIQL, and collecting data with the instrument as a whole. We consider the most recent version of the PIQL to be stable because qualitative and quantitative data analyses provide strong evidence for its validity as a tool for measuring students' PQL (see Sec.\ \ref{sec:validation}).
    }
    \label{fig:flowchart}
\end{figure*}

Development of the PIQL rested on two foundations: 1) existing literature in discipline-based education research in mathematics and physics, and 2) our own research on student reasoning about ratio, about signs and signed quantities, and about covariation. The latter was conducted over an approximately five year period immediately preceding our explicit development of PIQL as a standardized assessment instrument. Many PIQL items have been pilot-tested over the past decade as interview prompts, free-response written questions, and/or multiple-choice questions \cite{brahmia2015,boudreaux2015,brahmia2016exploring,brahmia2021PITs}. As we drew on these intellectual foundations, we followed an iterative cycle that led to the current version of the PIQL (see Fig.~\ref{fig:flowchart}). This cycle consisted of: assembling a working inventory of multiple-choice items, gathering data from introductory physics students, analyzing the results, identifying areas where the inventory could be improved (in terms of broader content coverage, items that were either too easy or too hard, redundant items, etc.), developing new items, adding and removing items from the working inventory and repeating data collection and analysis. %The cycle is represented in Fig.~\ref{fig:flowchart}. 

Our work was guided in a general way by the four-phase process for developing assessment instruments proposed by Adams and Wieman \cite{adams2010}:
\begin{itemize}
    \item Phase 1: Delineation of the purpose of the test and the scope of the construct or the extent of the domain to be measured;
   \item Phase 2. Development and evaluation of the test specifications;
    \item Phase 3. Development, field testing, evaluation, and selection of the items and scoring guides and procedures; and 
    \item Phase 4. Assembly and evaluation of the test for operational use.
\end{itemize}

In particular, the process described by Adams and Wieman informed the collection of both qualitative and quantitative evidence for validity of the PIQL. 

%Below, in part \ref{ssec:quantAndModeling}, we describe how we operationalize physics quantitative literacy and how PQL shapes both the content and form of PIQL test items. In part \ref{ssec:inventoryDev}, we discuss choices we made when assembling versions of the PIQL. In part \ref{ssec:itemDev}, we discuss our process for developing individual items. Finally, in part \ref{ssec:administration}, we describe the collection of data used to guide revisions to the PIQL, as well as the development of scalable and sustainable administration techniques.

\subsection{Operationalizing PQL}
\label{ssec:quantAndModeling}
%introduction to facets
Introductory physics courses present many new and abstract quantities, most of which are ratios, products, sums, or differences of other quantities. Quantities have associated units, and many can be positive or negative, where the sign carries physical meaning (e.g., negative work, positive charge). Quantities can also be vectors or scalars---which have different algebraic rules. Beyond additional meanings specific to the physical context, each of these aspects of quantity has associated \emph{mathematical} reasoning that is rich, nuanced, and challenging, as evidenced by research in mathematics education \cite{thompson2010,vlassis2004,thompson2003}.

Introductory physics primarily involves basic mathematics, such as arithmetic and algebra. A proceptual understanding of this mathematics is needed to support creative, quantitative reasoning with unfamiliar quantities. PQL thus involves using familiar mathematics in idiosyncratic ways, as exemplified in the spring example in the introduction. The PIQL was designed using three facets of quantitative modeling that are at the heart of quantification in introductory physics: proportional reasoning, reasoning with signed quantities, and covariational reasoning \cite{thompson2014,thompson2010,thompson2003,Sherin2001}.  PIQL items involve reasoning that draws on a nuanced conceptual blend of mathematics and physics content within each facet. We emphasize that the items on the PIQL focus on quantitative reasoning rather than procedural mathematics. While there are specific procedural topics that are not covered (e.g., operations with vectors in 2 or 3 dimensions, or explicit differentiation or integration), many PIQL items probe students' understanding of the underlying conceptual mathematics associated with mathematical procedures (e.g., the meaning of a negative sign associated with a vector quantity, or interpretation of a slope or area-under-a-curve). We believe the conceptual mathematics and quantitative reasoning covered by PIQL items are a foundation for the majority of mathematics used in college-level introductory physics courses.

%prop reasoning
\subsubsection{Facet 1: Proportional reasoning}
\label{sssec:pr}

The use of ratios and proportions to describe systems and characterize phenomena is a hallmark of expertise in STEM fields, perhaps especially in physics. Boudreaux, Kanim, and White Brahmia developed a set of six proportional reasoning subskills, based on their analysis of college students' specific difficulties on proportional reasoning assessment items \cite{boudreaux2015}. The subskills are consistent with the early work of Arnold Arons, who identified ``underpinnings'' to success in introductory physics, such as facility with the operational definitions that underlie familiar mathematical formulas (e.g., area~=~length $\times$ width), the verbal interpretation of ratios, including the slope of a graph, and reasoning with functional relationships (e.g., ``if the distance between two point charges doubles, the force decreases by one-quarter'') \cite{arons1983}. Three of the subskills---interpreting, applying, and constructing ratios---are particularly important in the PIQL \cite{Boudreaux2020perc}.

\subsubsection{Facet 2: Reasoning about sign}
\label{sssec:ned}
Negative pure numbers present more cognitive challenge than positive pure numbers for pre-college mathematics students \cite{bishop2014}. Mathematics education researchers have isolated a variety of ``natures of negativity'' fundamental to algebraic reasoning in the context of high school algebra---the many meanings of the negative sign that successful students distinguish and understand \cite{gallardo1994,thompson1988,nunes1993}. These various meanings of the negative sign form a foundation for scientific quantification, where the mathematical properties of negative numbers are well suited to represent natural processes and quantities. Physics education researchers report that a majority of students enrolled in a calculus-based physics course struggled to make meaning of positive and negative quantities in spite of completing Calculus I and more advanced courses in mathematics \cite{brahmia2016obstacles,brahmia2021PITs}. Developing ``flexibility'' with negative numbers, i.e., the recognition and correct interpretation of the multiple meanings of the negative sign, is a known challenge in mathematics education. There is mounting evidence that reasoning about negative quantity poses a significant hurdle for physics students at the introductory level and beyond \cite{White2020neg,Hayes2010,bajracharya2012,huynh2018,eriksson2018,ceuppens2019}.

%covar
\subsubsection{Facet 3: Covariational reasoning}
\label{sssec:cr}

Covariational reasoning, i.e., the formal reasoning about how one variable changes due to small changes in another, related quantity, has been shown by mathematics education researchers to be strongly associated with student success in calculus \cite{Saldanha1998, Thompson1994}. Early PER encompassed non-linear scaling and functional reasoning within proportional reasoning, which we re-contexualize here using the language of covariational reasoning established by mathematics education researchers \cite{Carlson2002, Moore2013, oehrtman2008}. Covariational reasoning encompasses all functions that relate two or more quantities and considers the multifaceted ways one can think about those relationships. Through this lens, proportional reasoning could be considered a subset of covariational reasoning that focuses on linear relationships. Considering its prevalence in introductory physics, proportional reasoning will be treated as a separate facet of quantitative literacy in this work. Physics education research is only beginning to explore how covariational reasoning is used in introductory physics contexts. Preliminary work suggests that covariational reasoning of physics graduate students (``experts'' in introductory physics contexts) differs in some ways from that of mathematics graduate students \cite{Zimmerman2019perc}.

%Together, proportional reasoning, reasoning with signs, and covariational reasoning form an essential core of quantitative reasoning important in introductory physics. While there may be other important sub-domains of PQL, we have taken these three as a foundational due to their importance in quantification and quantitative modeling in college-level introductory physics courses. 

\subsection{Assessing Reasoning Blends}

We consider the reasoning associated with each of our facets of PQL to be a product of well-formed \textit{conceptual blends} \cite{Fauconnier2002}. In order to assess the reasoning engendered by these blended reasoning spaces, the item library includes several ``multiple-choice/multiple-response'' (MCMR) items. Students are told that these items may have more than one correct response, and encouraged to choose all responses they believe are correct. An example MCMR item that we include on the PIQL is shown in Fig.~\ref{fig:workProb}. Complete understanding is demonstrated by selecting both D and G. In one study using an open-ended version of this question, White Brahmia observed that students were more accurate in describing the system energy (i.e., equivalent to selecting choice G in Fig. \ref{fig:workProb}) if they also mentioned a statement similar to choice D \cite{brahmia2021PITs}. A proceptually sophisticated response involves making sense of the negative sign by reasoning about both the orientation of the vector quantities as well as the physical ramifications of these vectors having opposite directions. We hypothesize that it is in the blended space that the reasoning becomes expert-like.

\begin{figure}[tb]
\framebox{\parbox{0.45\textwidth}{\raggedright
A hand exerts a constant, horizontal force on a block as the block moves along a frictionless, horizontal surface. No other objects do work on the block. For a particular interval of the motion, the hand does $W= -2.7$~J of work on the block. Recall that for a constant force, $W = \vec{F}\cdot\Delta\vec{s}$.\\

\vspace{1em}

\noindent
Consider the following statements about this situation.  Select the statement(s) that \textbf{must be true.}\\ \textit{\textbf{Choose all that apply.}}

\vspace{1em}

\begin{itemize}
\item[A.] The work done by the hand is in the negative direction. 
\item[B.] The force exerted by the hand is in the negative direction. 
\item[C.] The displacement of the block is in the negative direction. 
\item[D.] The force exerted by the hand is in the direction \textit{opposite to} the block's displacement. 
\item[E.] The force exerted by the hand is in the direction \textit{parallel to} the block's displacement. 
\item[F.] Energy was added to the block system. 
\item[G.] Energy was taken away from the block system. 
\end{itemize}

}}

    \caption{PIQL MCMR item that exemplifies a proceptual understanding of the negative sign in a mathematics and physics blend. The correct responses are D and G.}
    
    \label{fig:workProb}
\end{figure}

\subsection{Instrument Development}
\label{ssec:inventoryDev}
%Phases 1 and 2 (brief)
%As described above in Sec.~\ref{ssec:quantAndModeling}, the PIQL is intended to probe proceptual algebraic reasoning in contexts relevant to introductory physics, i.e., physics quantitative literacy. By collaboratively reflecting on our combined experience as instructors, as well as reviewing work in both mathematics and physics education research, we identified three facets as foundational to PQL: proportional reasoning, reasoning about signs and signed quantities, and covariational reasoning. 

%The \textit{multiple-choice/multiple-response} format items, described above, are useful for revealing details of how PQL develops over time. These items probe multiple facets of student reasoning about a given context, allowing researchers to track the development of those facets within a population of students completing a sequence of physics courses.

%Phase 3: Assembly of items from PRAT and early negativity items, and PCA questions

Like other inventories in physics, we originally intended the final version of the PIQL to have 20--30 items, and to take 30--40 minutes for students to complete. The prototype version of the PIQL (``protoPIQL'') consisted of 18 items that focused primarily on two facets of PQL: ratios and proportions \cite{brahmia2015,brahmia2016exploring,Brahmia2017c}, and signs and signed quantities \cite{brahmia2016exploring,brahmia2017signed,White2020neg}. The protoPIQL also included two items on covariational reasoning taken with permission from the Precalculus Conceptual Assessment (PCA) \cite{carlson2010precalculus}. The right side of Fig.~\ref{fig:flowchart} shows our workflow cycle for creating and pilot-testing preliminary versions of the PIQL. 
Iterative revisions were made over several years that strengthen the evidence for the validity and reliability of the PIQL, reduce redundancies, and ensure that the three facets of PQL were all represented. Later versions of the PIQL include 20 or 21 items. Due to these iterative revisions, the items on the PIQL in each of the six data sets are slightly different; we label the data sets by their version of the PIQL: protoPIQL, v1.0, v1.1, v2.0, v2.1, and v2.2.

Quantitative evidence for the validity of the PIQL was obtained by administering each version to students in introductory physics courses at Institution 1 and performing various statistical analyses of the data (see Sec.~\ref{sec:quantval} for details). We used these results to identify items to remove from the PIQL and items to modify. %Factor analyses of student responses did not reveal an instrument structure well-aligned with our three identified facets of PQL, despite such structure being easy for physics experts to identify. We therefore relied on the reasoning frameworks described above as well as expert validation interviews to ensure that all facets of reasoning were represented on the PIQL.

Qualitative evidence for the validity of the PIQL was obtained through interviews with both physics students and expert physicists and mathematicians (see Sec.~\ref{sec:qualval} for details). Interviews with experts helped to identify the need to develop additional items to meet our desired coverage of all three facets of PQL.

\subsection{Item Development}
\label{ssec:itemDev}
\begin{figure*}
    \centering
    \includegraphics[width = 0.9\textwidth]{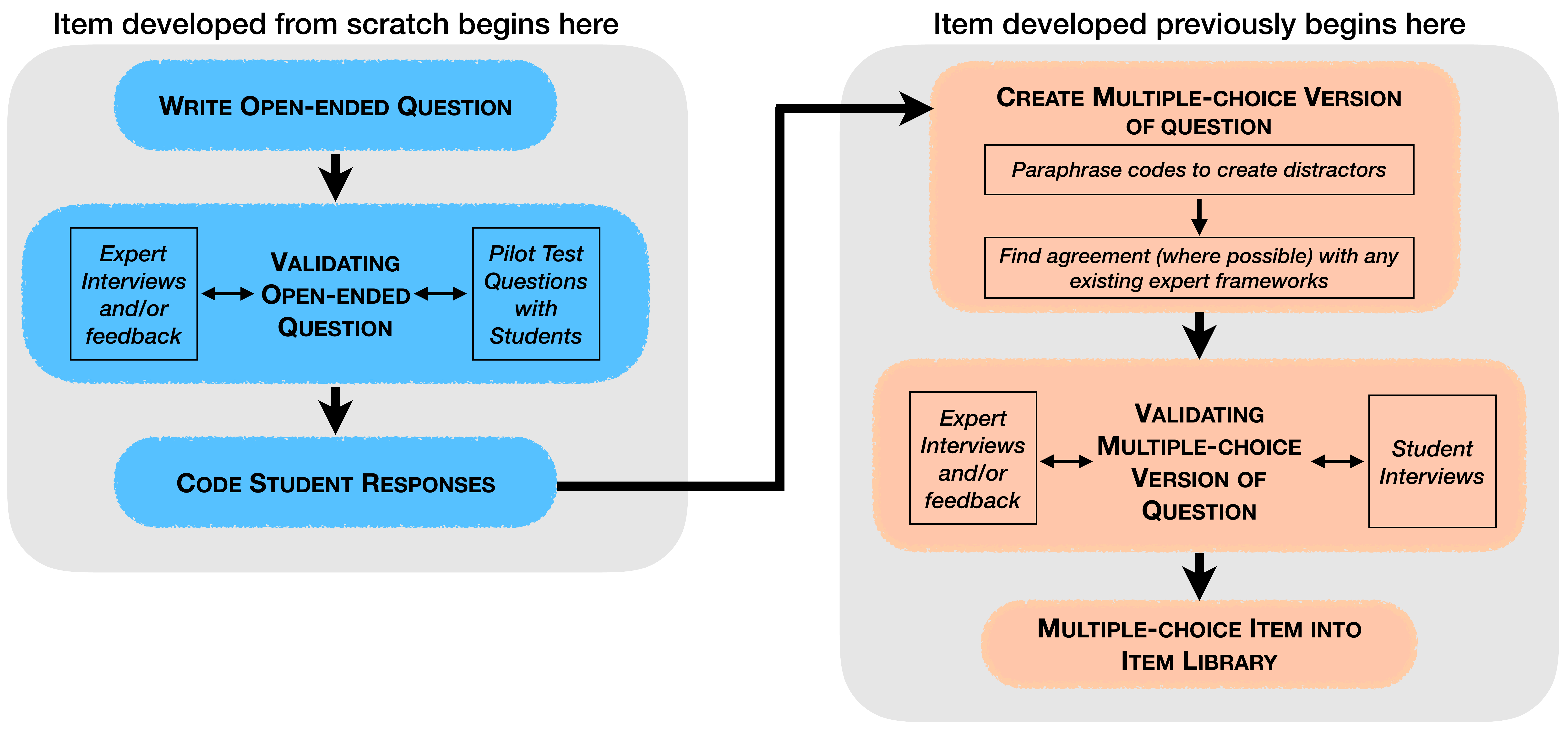}
    \caption{The initial development of items for the item library followed the processes shown above. The entire process applies to items that were developed from scratch. Items with existing evidence of validity, or those that had only a small finite number of possible answer options, enter the process at the top of the right side of the as indicated above. 
    }
    \label{fig:ItemDev}
\end{figure*}

% We identified the need to develop or modify items for the PIQL based on the results from administering versions of the PIQL in introductory physics courses and conducting panel interviews with physics and mathematics experts. 
Before creating the PIQL, we assembled a library of multiple-choice items that assess student reasoning in each of our three facets of PQL. Our item library originally contained items created during previous work of some of the authors and their collaborators \cite{boudreaux2015,Cohen2005,brahmia2014,brahmia2015,brahmia2016exploring,brahmia2021PITs}, and items from research on undergraduate mathematics education, including some from the Precalculus Concept Assessment (used with permission) \cite{carlson2010precalculus}. Fig.~\ref{fig:ItemDev} shows, in detail, our process for developing and modifying items to include in our item library. 

The need for developing new items was based on input from the interviews of physics and mathematics experts and our own judgement about the number of items involving each of our three facets of PQL. The need to modify items was often based on item-level statistical analyses of data collected from students at Institution 1.

% Items on the PIQL were generated in two ways. Approximately two-thirds of the items on the steady-state PIQL came from an existing library of questions. Many of the items in the existing library were developed in the course of our own collaborations investigating student quantitative reasoning \cite{boudreaux2015,Cohen2005,brahmia2014}. Other items in the baseline item library were adapted from questions developed by mathematics education researchers \cite{carlson2010precalculus} to assess covariational reasoning. The remaining items on the steady-state PIQL were developed specifically for inclusion on the PIQL. 

Items developed for the PIQL focus on reasoning rather than computation skill; most items require neither a calculator nor significant mental computation. %Here we describe how items were generated and revised. 
% In contrast to items developed as part of our prior research, some of the items on the steady-state PIQL were developed later, as we assembled the instrument, to ensure that all three facets of PQL would be well-represented. 
Most of these items involve reasoning about signed quantities and about the covariation of quantities. Some items were developed ``from scratch'' and do not have a basis in earlier items; others grew out of an iterative process of item modification.

Most items chosen for the steady-state PIQL---both those that grew out of prior work and those developed specifically for the PIQL---have undergone changes over the course of administration. Modifications of the items took place through an iterative cycle of research, item administration, and item validation. Most modifications were rooted in our deepening understanding of aspects of quantification and quantitative modeling in introductory physics. Generally speaking, modifications to question stems were relatively minor; however, we did make significant changes to many answer choices to encompass not only observed patterns in student reasoning but also multiple natures of expert reasoning. For example, on questions probing student reasoning about the negative sign in a given context, we used results from earlier investigations of student reasoning about negativity to develop item distractors. The wording of these distractors was modified while retaining the student ideas to be consistent with the meanings of the negative sign described by the framework of the natures of negativity of introductory physics. This framework categorizes various meanings and interpretations of the negative sign in introductory physics contexts \cite{White2020neg}; its use as part of our process is shown on the right hand side of Fig.~\ref{fig:ItemDev}. These distractors were then tested via student and expert interviews to ensure that the correct responses were consistent with expert reasoning and that the distractors were incorrect but consistent not only with common student reasoning but also with expert reasoning that might be correct in other contexts. Student validation interviews also led to improvements in the clarity and purpose of items. For further discussion of the validation process, see Sec.~\ref{sec:validation}.

As an example of an item that underwent substantial iteration, we describe the development of the ``Charged Spheres Question,'' which assesses student interpretation of the negative sign in the context of electric charge. Electric charge involves an idiosyncratic use of sign with quantity in physics---in this context, the sign is an indication of \emph{type} rather than an indication that the quantity is mathematically negative or ``less than zero'' \cite{olsho2021a}. An item developed as a part of work investigating student understanding of negativity \cite{brahmia2016exploring} in the context of the transfer of charge from one object to another was administered on the protoPIQL. When expert validation interviews with this item revealed ambiguities that could not be eliminated through item modification, we began to develop a new item involving the transfer of electric charge. The new item was developed as MCMR, to assess student understanding of the meaning of the negative sign in the context of electric charge, as well as identification of electrons as carriers of charge. Analyses of student responses indicated that the focus on charge carriers was unnecessary. We rewrote the item to focus on the meaning of the negative sign, and changed the wording to reduce ambiguity; the narrowed focus resulted in fewer answer choices, and we were able to rewrite the item as a single-response item. Although the new item shares a physics context (i.e., transfer of electric charge from one item to another) with the original item, the surface features of the current version of the item are distinct. Expert interviews indicate that the final version of the item, shown in Fig. \ref{fig:negSpheresMCSR}, is well-aligned with expert understanding of the meaning of the negative sign in the context of electric charge. With these changes, the Classical Test Theory (CTT) statistics for the item fall within the desired ranges (see section \ref{sec:validation} for more info on quantitative validation of assessment items).

\begin{figure}[tb]
\framebox{\parbox{0.45\textwidth}{\raggedright
A student has two electrically neutral aluminum spheres, A and B. The student performs an experiment that causes charge to move from one of the spheres to the other. After the experiment, the charge of sphere A is $q_{\textrm{A}}=-5$~microcoulombs, and the charge of sphere B is $q_{\textrm{B}}=+5~$~microcoulombs.\\

\noindent
What are the meanings of the negative and positive signs in this context?

\begin{itemize}
    \item[I] The signs imply that the charge on sphere A is less than that on sphere B.
    \item[II] The signs imply that the unbalanced charges on the two spheres are of opposite types.
    \item[III] The signs imply that charge was removed from sphere A and added to sphere B.
\end{itemize}

\vspace{1em}

\begin{itemize}
\item[A.] I
\item[B.] II
\item[C.] III
\item[D.] I and II
\item[E.] I and III
\item[F.] II and III
\item[G.] I, II, and III
\end{itemize}
}}

    \caption{Current version of the \textit{Charged Spheres} item on the steady-state PIQL, which has withstood multiple challenges to its validity, intended to assess student reasoning about the negative sign in the context of electric charge. Answer B is correct.
    }
    \label{fig:negSpheresMCSR}
\end{figure}

\subsection{Administration of the PIQL during development}
\label{ssec:administration}
We had two primary goals related to administration of the PIQL during its development: 1) collection of data from students to construct an argument for the validity of the PIQL as a whole test and individual PIQL items; and 2) development of methods for the administration of the PIQL that are scalable and sustainable, and improve its utility for physics instructors and researchers. %In this section, we describe our progress toward each of these goals.

During initial development we administered the PIQL to all students enrolled in the 3-quarter, large-enrollment, calculus-based introductory physics sequence at Institution 1. We ran versions of the PIQL over eight academic quarters. It was administered at the beginning of each term, before significant instruction, thus serving as a ``pretest'' for each course of the introductory sequence. Consistent with best practices, the PIQL was originally designed to be administered on paper and in-person (proctored) \cite{madsen2017tpt}; however, administration protocols were modified in response to the COVID-19 pandemic. Reference \cite{Olsho2020perc} contains a detailed description of both our in-person and online administration methods.

For most in-person administrations of the PIQL, students read items from a 5-page stapled packet and recorded their responses on a paper answer form as well as electronically, where the paper copy served as a backup if the electronic submission failed. Because students entered their responses electronically, we were able to download data in a format that allowed for statistical analyses without substantial cleaning. We developed this hybrid paper/electronic data collection system because the inclusion of several ``multiple-choice/multiple response'' (MCMR) items on the PIQL prevented the use of Institution 1's existing system for scanning multiple-choice responses from bubble sheets. % Some students were unable to enter their answers online, so a member of the research team entered those responses manually. Responses for roughly 25--50 students were added manually each quarter. %As is common with large-enrollment courses, some students misunderstood how to submit their responses to the PIQL, resulting in research team members spending additional to time make sure the data set was complete and that students received credit for their work. 

Though our focus was on in-person, proctored administration of the assessment, we began to consider whether online, unproctored administration would better support broader validation and wide-spread dissemination. While existing research suggests little or no significant difference in student performance between proctored and unproctored administrations of some research-based assessments \cite{bonham2008,nissen2018,wilcox2019prper}, researchers recommend that online, unproctored administration be validated separately \cite{bonham2008}. We wanted to determine whether the PIQL could be administered online and unproctored by instructors who were reluctant or unable to allocate class time for administration. %Moreover, though we generally have access to students during the first week of classes during scheduled recitation sessions, scheduling was difficult during academic quarters in which instruction started midweek, leading to confusion and decreased participation rates.

The COVID-19 pandemic forced the issue. With both Institution 1 and Institution 2 moving to all online instruction over a very short time interval, in-person administration of the assessment became impossible. Although online ``proctoring'' services exist \footnote{Many of these systems can be described more accurately as surveillance---they cannot interact with the students or provide a physical presence.}, the proctoring requirements do not align well with Institution 1's policies regarding computer camera use during virtual instruction, and do not take into account possible limitations on students during such an uncertain and difficult period. 

We ran the PIQL unproctored and entirely online using Institution 1's existing survey/quiz platform. To mitigate student stress during the rapid shift to online learning, Institution 1 suggested that no graded work be required during the first week of instruction. Because we do not grade students' responses to the PIQL for correctness, we decided to run the PIQL, as usual, during the first week of the term in each of the three courses of the calculus-based introductory physics sequence. As we were aware of the tendency of some students to place undue importance on such assessments, however, we presented the PIQL as a low-stakes survey. 

We adhered to best practices for online administration of a low-stakes assessment instrument \cite{wilcox2018,nissen2018} as much as possible: we used a time limit substantially longer than the amount of time expected to complete the assessment; we embedded a short (less than three-minute) video at the beginning of the online PIQL that explained the purpose of the assessment and emphasized that course credit would be awarded based on completion rather than correctness; we sent students multiple reminders to complete the PIQL to improve overall participation rate; and we constructed the online version of the PIQL to discourage copying or saving of test items \cite{Olsho2020perc}. Little prior work has been done to establish best-practice recommendations regarding the administration of MCMR items in an online format. While Wilcox and Pollock report that student performance on coupled multiple-response items on the Colorado upper-division electrostatics (CUE) diagnostic was similar across online and in-person administration methods, an examination of possible differences was not a focus of the work \cite{wilcox2015}. We also note that the coupled multiple-response items on the CUE are a different style of question than the MCMR items on the PIQL, and further, that the CUE and the PIQL are used with different populations of students (upper-division vs. introductory level). %Below we describe our approach to and experience with administering MCMR items online to students enrolled in introductory physics.

Six of the 20 items on the PIQL (v2.2) are MCMR. When the instrument was administered in person, there were multiple opportunities to remind students that they could choose more than one response on these items: both in writing on the instrument itself, and also verbally by the proctor. Validation interviews suggested that multiple reminders were necessary, as this variety of question is relatively rare on the assessments typically encountered by students. Because we recognized that a majority of students completing the survey for the first time would have little-to-no experience with MCMR items, we wanted to increase the likelihood that students would recognize that they could select multiple responses for those items when encountering them online (unproctored). All of the MCMR items were moved to the end of the survey for the online administration. After answering the last multiple-choice/single-response item, students saw a page containing only a statement that the remaining questions on the survey might have more than one correct response, and that students should choose all answers that they feel are correct. At the top of the page for each of the remaining items (all MCMR), students saw a reminder that the question might have more than one correct response. As with the paper version of the PIQL,we also prompted students to ``choose all that apply'' in the question stem for each MCMR item (see, for example, Fig.~\ref{fig:workProb}).

Response rates for the MCMR items suggest that the measures described above were effective; further, overall results from online administrations of the PIQL suggest that students take the online-version of the assessment seriously and perform at roughly the same level as for in-person administration \cite{Olsho2020perc}. This is consistent with work that indicates that low-stakes assessments can be successfully administered online \cite{bonham2008,nissen2018,wilcox2019prper}. We note that small but significant differences are seen in student performance on MCMR items \cite{Olsho2020perc}. A preliminary investigation of these differences suggest that the differences in performance on the MCMR items is due to the difference in administration method (i.e., in-person and on paper vs. online), and is described in a manuscript in preparation \cite{olsho2021mcmr}. 

%While we are not yet confident that results from in-person administration of the PIQL can be compared to those from an online administration
Our iterative process of continually revising the PIQL combined with the sudden need to administer it online (due to the COVID-19 pandemic) prevented us from collecting student responses on identical versions of the PIQL being administered both in-person and online \footnote{The first version of the PIQL administered online (v2.2) differs by onle one item from the final version administered in-person (v2.1).}; however, we have collected (and continue to collect) substantial evidence of the validity of the PIQL as an online instrument (see Sec.\ \ref{sec:validation}). Our ultimate goal is to disseminate the PIQL as an instrument that can be administered pre/post-instruction either online or in-person.

\section{Evidence for the Validity and Reliability of the PIQL}
\label{sec:validation}

One of the major efforts of our project involves establishing evidence for the validity of using the PIQL to measure students' PQL in introductory physics. We established qualitative evidence of validity by conducting interviews with physics experts and students. We established quantitative evidence of validity by administering the PIQL to students in the calculus-based introductory physics sequence and analyzing their responses using a variety of statistical and psychometric methods.

Our goals in establishing various pieces of evidence are to show that
\begin{itemize}
    \item students interpret individual PIQL items and response choices as intended;
    \item physics experts agree on the correctness of individual items, and view the PIQL as a whole as a valuable tool for use in introductory physics courses;
    \item PIQL items represent a breadth of difficulties that are neither unattainable by incoming physics students nor trivial for students completing the introductory sequence;
    \item student performance on individual PIQL items roughly correlates with their performance on the test as a whole;
    \item students choose incorrect responses (a.k.a.\ ``distractors'') with nontrivial frequency; and
    \item student responses do not indicate PIQL items are redundant with each other.
\end{itemize}

The actions taken to analyze data may be seen in the gray diamonds on the sides of Fig.\ \ref{fig:flowchart}. We used the results from these analyses to inform revisions to individual items, the development of additional items, and revisions to the PIQL as a whole in terms of selecting which items to include. These efforts have resulted in a stable instrument with strong qualitative and quantitative evidence for its validity as a measure of students' PQL.

\subsection{Qualitative evidence of validity}
\label{sec:qualval}
Face validity of the PIQL was assessed primarily through expert and student interviews. Interviews were used to collect evidence of the validity of the PIQL and individual PIQL items in three distinct ways:

\begin{enumerate}
	\item Expert panel reviews were performed to verify that the correct answer choices are indeed correct, as well as to ensure that distractors are incorrect; experts also identified the mathematical content of each question, which allowed us to assess the face validity of the individual items.
	\item Individual student interviews were performed to determine whether students are interpreting the questions as intended, to ensure that students are choosing the correct answer for the correct reasons, and to ensure that incorrect responses are chosen for consistent reasons.
	\item Expert validation interviews were performed to verify that individual items and the assembled inventory are testing ideas that experts expect their students to learn.
\end{enumerate}

During the expert panel reviews, physics education research faculty and graduate students worked in groups of 3--5 members, with each group seeing 6--8 questions. Panel members worked through each item individually to determine the correct answer and identify the specific mathematical construct(s) required to answer the question.  The panels then discussed the items as a group, to come to a consensus about the correct answer and that the incorrect answers were indeed incorrect; together, they also rated the quality of the question in terms of clarity, ambiguity, and appropriateness. Researchers observed the conversations, took notes, and collected the materials afterwards.

\begin{figure*}[t]
    \includegraphics[width = 0.8\columnwidth]{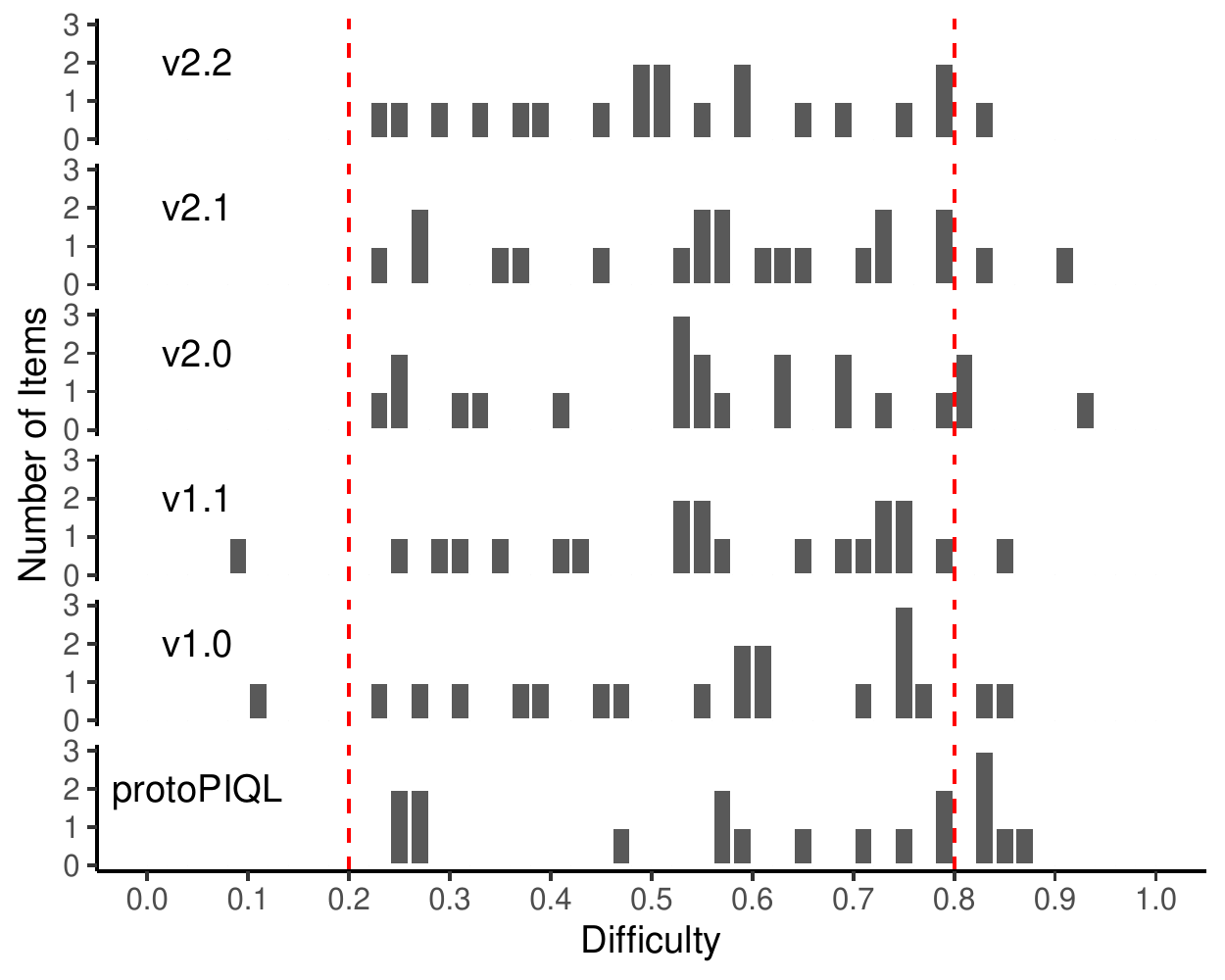}
    ~~~~~~~~
    \includegraphics[width = 0.8\columnwidth]{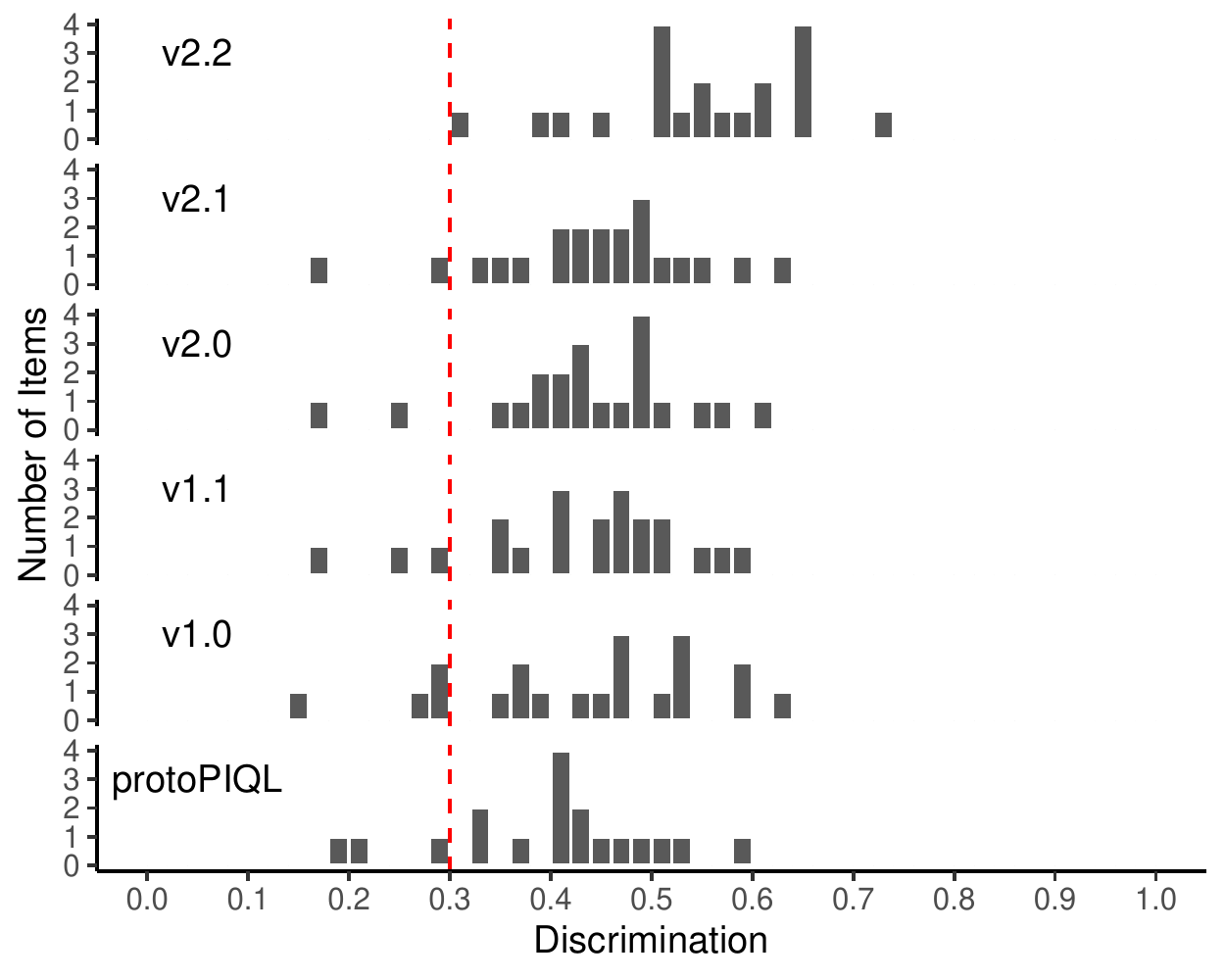}
    
    ~~~~~~(a) \hspace{0.83\columnwidth} (b)
    \caption{CTT difficulty (a) and discrimination (b) parameter distributions for all versions of the PIQL. Successive versions are ordered from bottom to top. The desired range of difficulty values is between 0.2 and 0.8 (shown by dashed red lines). The desired range for discrimination is above 0.3.}
    \label{fig:CTT}
\end{figure*}

For the individual student interviews, students were recruited in approximately equal numbers from each of the three courses in the calculus-based introductory physics sequence at Institution 1. During interviews lasting 30--60 minutes, the students were asked to work through the questions to be validated, following the same``think-aloud'' protocol used during the initial development of the assessment items, described above in Sec.\ \ref{ssec:itemDev}. The focus of these interviews was not to explore student reasoning about the material, but to ensure that students were interpreting the items as intended, and choosing the correct answer(s) for the right reasons.

A small number of interviews following this protocol were also performed with students enrolled in the calculus-based introductory physics sequence at Institution 2. The student validation interviews informed changes to the questions to improve their coherence with the target population. %Finally, we performed expert interviews for the complete PIQL, as described at the end of section \ref{ssec:inventoryDev}.

%For another round of expert interviews, a complete version of the PIQL was sent out to instructors with extensive experience teaching in the introductory physics sequence at Institution 1. The experts were asked to review the PIQL before the interviews. During formal, semi-structured interviews, experts were asked to comment on the appropriateness of the items and of the test as a whole to ensure that the PIQL is testing ideas that experts expect their students to learn. The interviews were recorded and the interviewers took written notes. These interviews resulted in a number of small but substantive wording changes for two of the items to improve their clarity. Feedback about the relevance of the items with respect to course learning objectives also informed the composition of the PIQL as a whole, resulting in one item being removed from the instrument. 

As we approached a steady-state version of the PIQL we performed validation interviews with both mathematics and physics experts using the PIQL as a whole. Our two mathematics experts are faculty members in the Departments of Mathematics at two different public universities, both with experience teaching college-level calculus. The validation interview with the mathematics experts was performed as a group interview with both experts attending virtually. Our seven physics experts are faculty members of the Department of Physics at Institution 1 and were chosen for their extensive experience teaching in the introductory physics sequences. We held an in-person group session for the validation interviews with our physics experts. Before both of the group interviews, the experts were asked to complete the PIQL and submit their responses to us. At both group interviews, we discussed each item to ensure that there was consensus for the correct responses and to check that the items were clearly worded. We also discussed in detail the reasoning each item addresses, and the appropriateness of the required reasoning for our target population of students. These expert interviews served as a final check of the individual items, and of the PIQL as a whole. Experts agreed that, overall, the PIQL represents reasoning that they expect their students to develop during introductory physics courses and that is important in physics generally. We removed one item experts felt did not represent reasoning central to introductory physics. Physics and mathematics experts agreed that administering the PIQL to their students would give important information about the students' quantitative reasoning, and how that reasoning changed over instruction. These results provide evidence for the validity of the PIQL as a whole.

\subsection{Quantitative evidence of validity using Classical Test Theory}
\label{sec:quantval}
We have previously published a detailed account of various quantitative analyses to measure the validity and reliability of each version of the PIQL \cite{Smith2020perc}. Here we present the highlights of this work, emphasizing evidence that the most recent version of the PIQL consistently meets our expectations for quantitative measures of validity and reliability. 

Using Classical Test Theory (CTT) we calculated the difficulty and discrimination parameters for each item; we want to have a wide range of difficulty values with most items between 0.2 and 0.8 (representing the fraction of students who answer each item correctly), and we want most discrimination values to be above 0.3 (representing the difference in CTT difficulty between the top and bottom 27\% of students) \cite{Wiersma1990}. We also calculated Cronbach's $\alpha$ as a measure of reliability: a value of at least 0.7 indicates that the test is reliable for measuring the performance of groups of students on a single-construct test, and a value of at least 0.8 indicates that the test is reliable for measuring the performance of individual students \cite{Doran1980}.

Fig. \ref{fig:CTT} shows the distribution histograms of the CTT difficulty and discrimination parameters for each version of the PIQL. One can see that, with each successive version (ordered from bottom to top in the figure), more items fall within the desired difficulty range (between the red dashed lines). Additionally, the discrimination values get progressively higher, with all 20 items in v2.2 being higher than the desired threshold (red dashed line). Additionally, Cronbach's $\alpha$ went from 0.67 on the protoPIQL to 0.80 on v2.2.

\begin{figure}[t]
    \centering
    \includegraphics[width = \columnwidth]{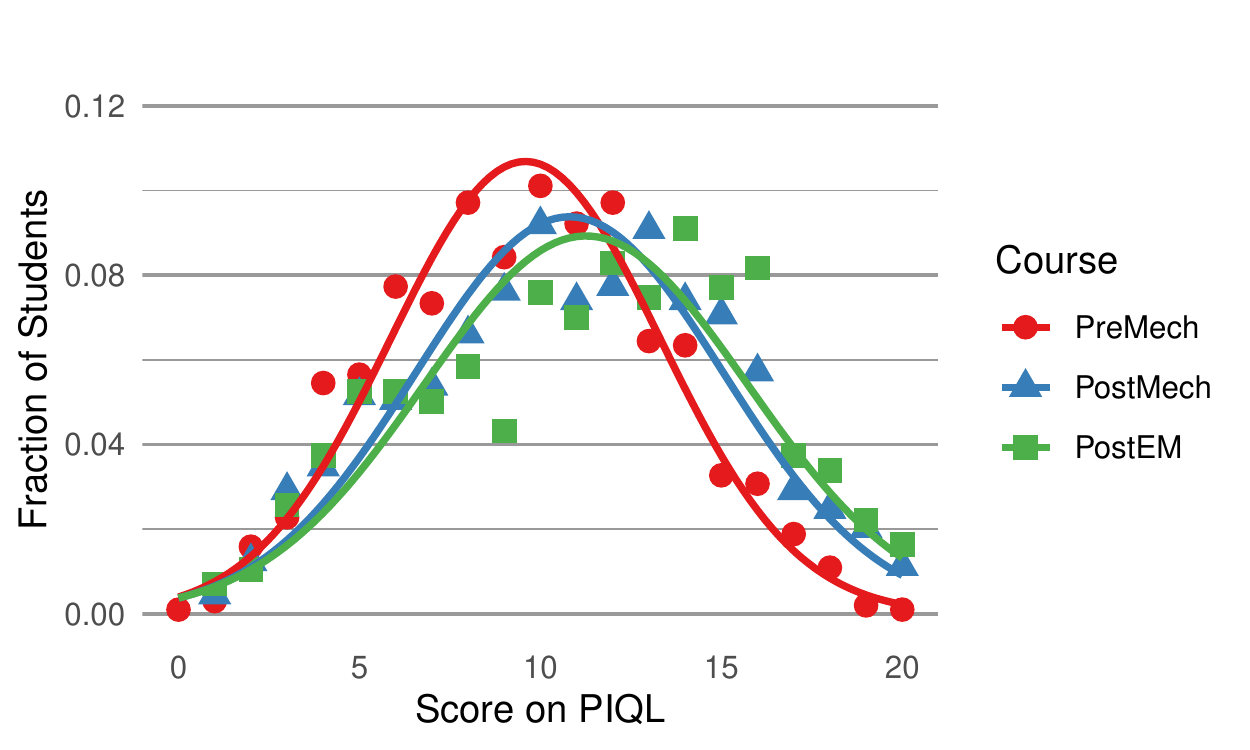}
    \caption{Distribution of PIQL scores for each course in which data were collected. Data include 2,758 response sets from 2,078 individuals who responded to PIQL v2.2 in four different academic terms. (Some students provided data in multiple courses at various times.)}
    \label{fig:scores}
\end{figure}

Fig. \ref{fig:scores} shows the score distribution for each course in which data were collected. As mentioned above, data were collected at the beginning of each course. We consider these to be three snapshots of the introductory course sequence: before mechanics ``PreMech'', after mechanics but before electricity \& magnetism ``PostMech'', and after electricity \& magnetism but before thermodynamics and waves ``PostEM''. The data in Fig.\ \ref{fig:scores} combine all four administrations of PIQL v2.2, including 2,758 data sets and 2,078 individuals \footnote{Because data from four academic terms are combined, some individuals provided multiple data sets at different times in their academic sequence.}. A one-way analysis of variance showed no significant difference between the four academic terms, $F(3, 2754) = 1.6$, $p = 0.2$.

Fig. \ref{fig:scores} shows that students entering the calculus-based introductory physics sequence at Institution 1 have moderate levels of PQL before instruction: the PreMech mean and standard deviation are $9.6 \pm 3.7$ out of 20. Students' PQL increases slightly with instruction, with mean scores $10.8 \pm 4.3$ for PostMech and $11.3 \pm 4.5$ for PostEM, but these score distributions show that many students have not achieved high levels of PQL after two introductory courses. Recent studies have shown that differences between student data from pre- and post-instruction research-based conceptual physics assessments typical result in a Cohen's $d$ effect size of about $d = 1$ \cite{Nissen2018a}. Effect sizes from PIQL data are significantly lower: $d = 0.3$ from PreMech to PostMech and $d = 0.1$ from PostMech to PostEM, despite data being collected in courses that make heavy use of research-based instructional materials. This suggests that PQL development, while an implicit goal of any introductory physics courses, may be limited without specific targeted instruction.

\subsection{Exploring the substructure of the PIQL}
\label{ssec:substructure}

The PIQL was initially developed to probe student reasoning about three facets of PQL that were defined from an expert's perspective: ratios and proportions, covariation, and signed quantities/negativity (see Sec.\ \ref{ssec:quantAndModeling}). In the language of factor analysis, this would imply that the PIQL was originally intended to elicit student responses that are consistent with a three-factor structure. Because the intended factor structure of the PIQL was well understood at the beginning of its development, confirmatory factor analysis (CFA) of student responses was used at the onset, in conjunction with exploratory factor analysis (EFA). CFA is a model-driven statistical method whose goal is to identify the adequacy of a proposed factor model for explaining patterns in response data from the instrument being analyzed \cite{Brown2015}. In our work, CFA helps reveal whether or not student response patterns align with the facet-driven model predicted by experts. EFA is a data-driven statistical method whose goal is to uncover the underlying dependencies between observed variables \cite{Lawley1963}. For all versions of the PIQL, CFA determined that the proposed, facet-driven, expert factor model was not an adequate representation of the latent trait structure of \textit{student} PIQL responses \cite{Smith2019perc,Smith2020perc}. The target threshold for CFA is to have goodness-of-fit statistics such as the Confirmatory Fit Index (CFI) and Tucker-Lewis Fit Index (TLI) above a threshold of 0.9 \cite{Eaton2018}. For all versions of the PIQL the CFI and TLI were below 0.8 when using the facet-driven factor model; therefore, students' response patterns did not match our expectations of reasoning developing differently in the areas of ratio and proportion, covariation, and signed quantities/negativity.

Given that the CFA results do not fit with the proposed factor model, we moved on to a more in-depth investigation using EFA. The goal of using EFA was to determine if the PIQL has any substructure, and how closely any substructure aligns with the three facets of PQL. EFA results from preliminary versions of the PIQL showed factor structures that did not align with our three facets of PQL. These results also allowed us to identify redundant items that loaded strongly onto a common factor, excluding all other items; we chose to eliminate redundant items to produce a more efficient instrument. 

Analyses of student responses to the most recent versions of the PIQL (v2.0, v2.1, and v2.2) suggest unidimensionality, with no strong substructure amongst the items \cite{Smith2020perc}. Results from EFA parallel analysis  and CFA suggest that these versions may be adequately described by a single factor. Goodness-of-fit analyses show that the CFI and TLI were above 0.93 for both versions under CFA using a unidimensional model. Additionally, the standardized root mean square of the residuals (RMSR) and the root mean square of the error of approximation (RMSEA) were both below 0.04, with a model being considered an adequate representation of the data with values below 0.06 and 0.08, respectively. Collectively these results show strong support for the claim that students answer PIQL items in a manner consistent with a unidimensional model \cite{Eaton2018}. We consider this to be evidence that PQL is a somewhat coherent construct for students that does not cleanly separate into various facets in the ways expected by experts.

\section{Analyzing Data from Multiple-Choice/ Multiple-Response Items}
\label{sec:mcmr}
A unique feature of the PIQL is our inclusion of multiple-choice/multiple-response (MCMR) items in which students are instructed to ``select all statements that \textbf{must be true}'' from a given list, and to ``\textbf{\textit{choose all that apply}}'' (emphasis in the original text). As described in Sec.\ \ref{sec:foundations} we consider PQL to be a conceptual blend between physics concepts and mathematical reasoning \cite{Fauconnier2002,White2020rume}. The MCMR item format has the potential to reveal more information about students' thinking than standard single-response items, and to measure the complexity of ideas that students bring from both of these input spaces \footnote{We note that items 4 and 14 on the PIQL  are single-response items that could be written as MCMR; these items had few enough answer choices that we could predefine combinations and use a single-response format.}. The MCMR item format also poses challenges with data analysis, as typical analyses of multiple-choice tests (such as CTT, EFA, and CFA) assume single-response items.

\begin{figure}[t]
    % \vspace{-6mm}
    \includegraphics[width = 0.45\textwidth]{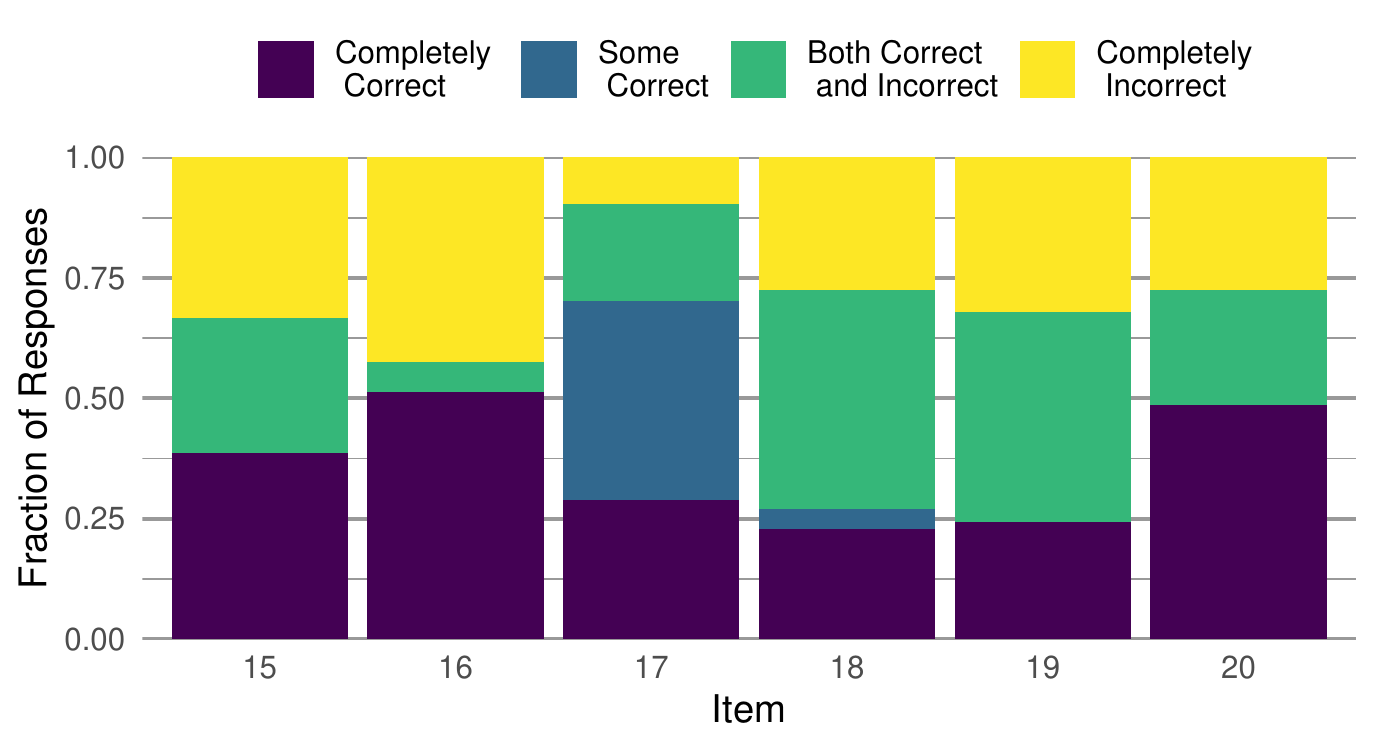}
    \caption{Fraction of student responses in each category of our four-level scoring scheme for MCMR items with multiple correct answers. These results are from PIQL v2.2.}
    
    \vspace{-3mm}
    \label{fig:correctness}
\end{figure}

% \subsubsection{Four-level scoring scale} 
For MCMR items, dichotomous scoring methods require a student to choose \emph{all} correct responses and \emph{only} correct responses to be considered correct. For example, item 18 on PIQL v2.2 has two correct answer choices: D and G (see Fig.~\ref{fig:workProb}). In a dichotomous scoring scheme a student who picks only answer D would be scored the same way as a student who chooses answers E and F (incorrect). This ignores the nuance and complexity of students' response patterns within (and between) items. As such, the CTT results for these items are not entirely representative of students' responses.

\begin{figure*}[bt]
    \centering
    \includegraphics[width = 0.3\textwidth]{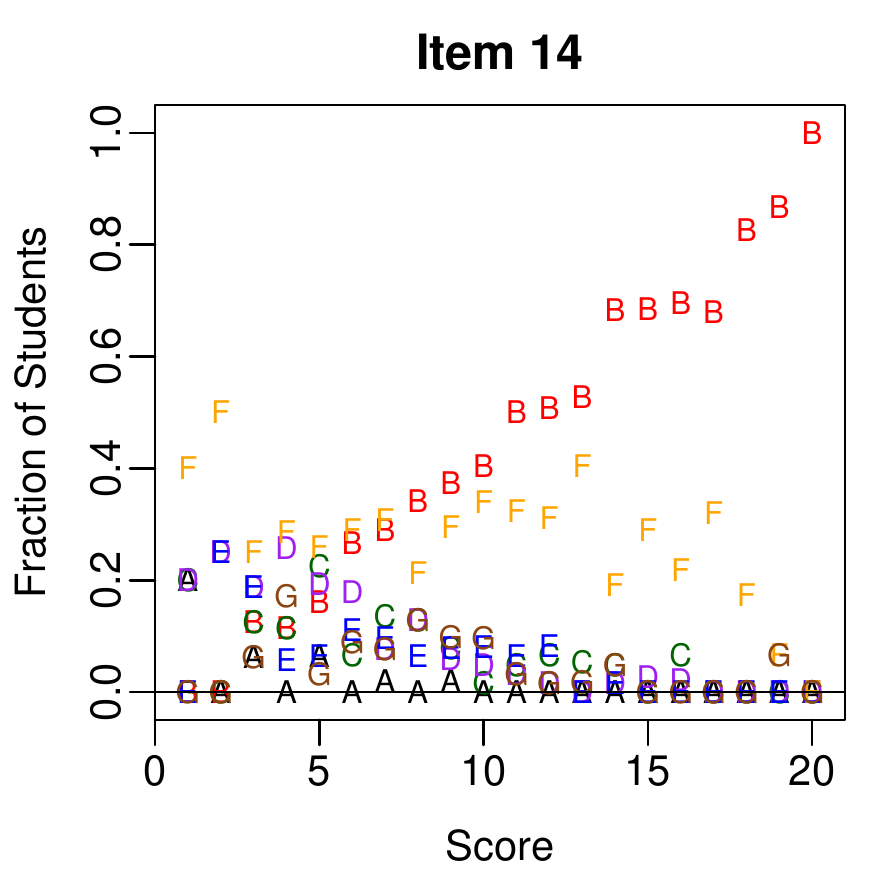} ~~~~ \includegraphics[width = 0.3\textwidth]{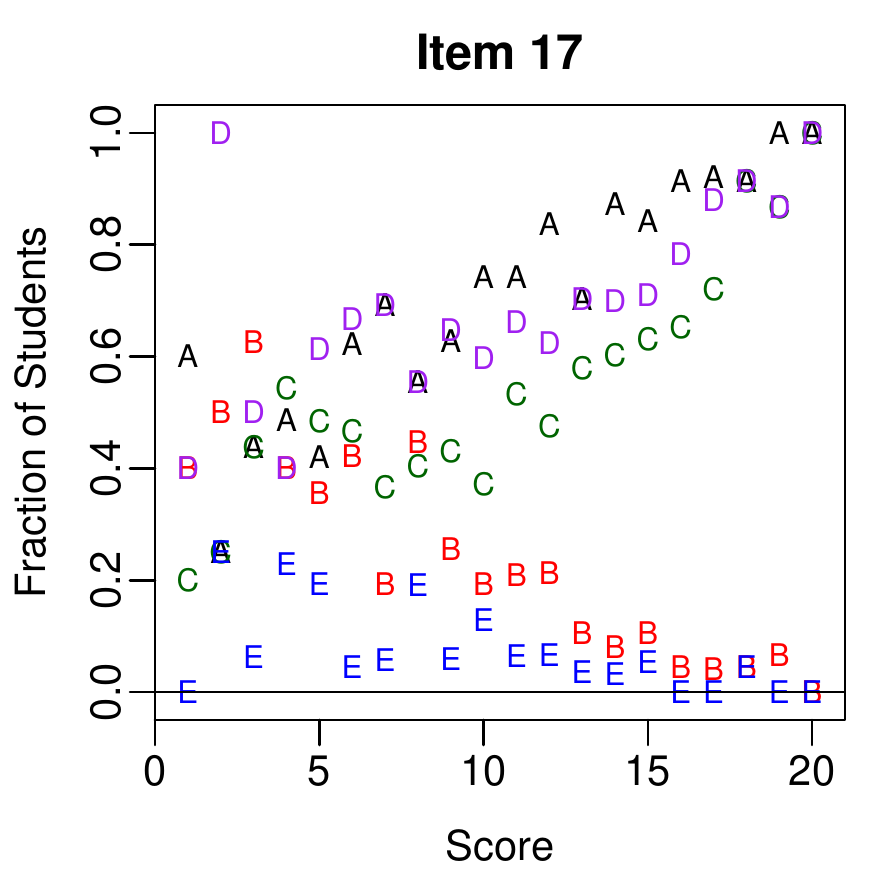} ~~~~ \includegraphics[width = 0.3\textwidth]{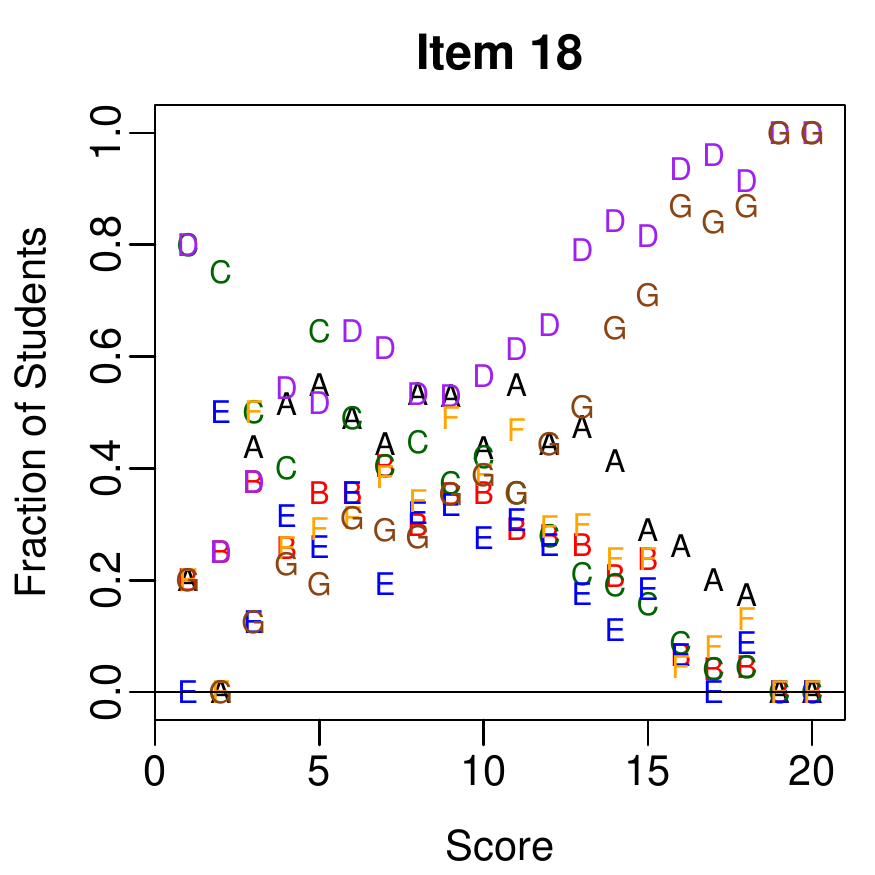}
    \caption{Item Response Curves for three items on PIQL v2.2. Each plot shows the fraction of students who chose each response out of the students who earned each score on the total test. Item 14 has correct answer B, item 17 has correct answers A, C, and D, and item 18 has correct answers D and G.}
    \label{fig:ircs}
\end{figure*}

In an effort to move beyond the constraints of dichotomous scoring for MCMR items, we have developed a four-level scoring scale in which we categorize students' responses as Completely Correct, Some Correct (if at least one but not all correct response choices are chosen), Both Correct and Incorrect (if at least one correct and one incorrect response choices are chosen), and Completely Incorrect \cite{Smith2018, Smith2019rume}. Fig. \ref{fig:correctness} shows the results of using this four-level scoring scale to categorize student responses to the six MCMR items on PIQL v2.2. The dark purple Completely Correct bars are equivalent to CTT difficulty; however, Fig.\ \ref{fig:correctness} also shows us that at least 60\% of students provide at least one correct response to each item (Completely Correct, Some Correct, and Both Correct and Incorrect combined), although this is often coupled with an incorrect response (6\%--45\% of students categorized as Both Correct and Incorrect). This tells a very different story than the CTT results, which group the Some Correct, Both Correct and Incorrect, and Completely Incorrect categories together into a broad Incorrect category.

These four-level scoring results also reveal differences hidden by dichotomous scoring. For example, on PIQL v2.2 two items (Items 17 and 18) have more than one correct answer choice. Fig. \ref{fig:correctness} shows that roughly equal numbers of students answers these items completely correctly, but item 17 has a much higher fraction of students in the Some Correct category. Students are much more likely to include one of the incorrect responses to item 18 than they are for item 17. Items with multiple correct answers also present a new question: is it better for a student to choose Some Correct answers or Both Correct and Incorrect answers? The answer may depend on the specifics of each item and the associated answer choices. Future work will include analyzing data from MCMR items to develop a more sophisticated scoring scheme.

% \subsubsection{Item Response Curves} 
% \label{ssec:irc}

To further examine the responses students give to individual PIQL items we use Item Response Curves (IRCs), which show the fraction of students who choose each answer choice as a function of the students' overall score on the PIQL \cite{Morris2006,Morris2012,Walter2016,Ishimoto2017}. IRCs have been used with single-response tests to rank incorrect responses and to compare different student populations with regard to both correct and incorrect answer choices \cite{Walter2016,Ishimoto2017}. We find IRCs particularly helpful for examining student responses to items with multiple correct answers. 

Fig. \ref{fig:ircs} shows three IRCs with different behavior. Item 14 is a single-response item with correct answer B. Even fairly high-scoring students persist in choosing a particular incorrect answer F \footnote{For the discussion about using IRCs we have purposely omitted the details of the context of items 14, 17, and 18. We want to highlight the use of IRCs to help identify different trends in student answers and see how these different between single-response and multiple-response items. A detailed discussion of why students answer these particular items in these particular ways is beyond the scope of this paper.}. Item 17 has three correct responses (A, C, D), with A being the most commonly chosen, and C being the least commonly chosen. Few students at any score level choose E, and fewer than 20\% of students who score above average (10.8) choose either incorrect response (B, E). Item 18 is particularly interesting in that all responses are chosen by 20\%--60\% of students in the middle score range (8--12), which represents 40\% of the students (see Fig.~\ref{fig:workProb} for the text of item 18 and Fig.\ \ref{fig:scores} for the score distribution). This supports the results from Fig.~\ref{fig:correctness} that students are likely to choose both a correct and an incorrect response to item 18.

Both the four-level scoring scheme and the IRCs provide more information than traditional CTT analyses and allow us to see patterns in students' responses that go beyond typical dichotomous scoring methods. We have used these to gain a deeper qualitative picture of student performance on each PIQL item, and these have been very valuable for deciding which items and answer choices to keep, eliminate, or modify. The MCMR item format also opens possibilities for future work to develop rigorous quantitative analyses that can supplement traditional analyses and provide richer information about student understanding and learning.

\section{Conclusion}
\label{sec:future}

We have developed the Physics Inventory of Quantitative Literacy (PIQL), an instrument to assess quantitative reasoning in introductory physics contexts. Instruction in introductory physics includes a greater amount and a deeper level of quantitative reasoning than introductory courses in most other STEM disciplines. We have presented expert interview data confirming that the promotion of such reasoning is an important goal of introductory physics courses, which are required for most STEM majors. 

We have described the development of the PIQL and presented evidence of its validity. We demonstrated that experts believe that the PIQL is a valid assessment of PQL. Results from quantitative analyses show strong evidence for the validity and reliability of the PIQL at Institution 1, and suggest that student responses align with a coherent unidimensional model of PQL. We have also provided evidence that students' PIQL scores don't improve much even after having taken two quarters of calculus-based physics; PQL is not a byproduct of current common research-based instructional practices. 

We have supplemented rigorous psychometric analyses with four-level scoring methods for MCMR items and IRCs, which provide additional information about students' choices of both correct and incorrect responses. These analyses informed our revisions of the PIQL. Future work will include developing more sophisticated analyses that can include the nuance of MCMR data into CTT-style analyses.

We continue to collect evidence for validation of the PIQL for use as an online reasoning inventory. We recognize that to construct a more compelling argument for the validity of comparing paper and online versions of the PIQL, we must learn more about how students interact with MCMR test items when using a computer or other internet-capable device. To this end, we intend to develop interview protocols to investigate how students interact with the PIQL, especially its MCMR items, when it is administered online. A more in-depth analysis of student performance on MCMR items for different administration methods is in progress.

Next steps involve collecting data to construct an argument of the PIQL's validity to assess PQL across more diverse student populations. Historically, physics education research studies oversample \cite{kanim2020} from large research universities that are similar to Institutions 1 and 2, where the work we describe in this paper has been done. The broader population of introductory physics students is a more racially and socioeconomically diverse group, attending a variety of geographically diverse post-secondary institutions \cite{kanim2020}. 

The use of the PIQL can raise awareness in undergraduate physics education, and STEM education more broadly, across varied post-secondary learning environments. We believe that a broader awareness has the potential to inform instructional practices and help drive change that will deepen student PQL. It is essential that materials and methods developed with a relatively homogeneous student sample be validated for the breadth of learning environments in which students take introductory physics. A PIQL with broader evidence of validity will serve as a new tool for improving the ability of instruction to promote quantitative literacy.

We also seek to use the PIQL as a tool for characterizing the \textit{development} of students' PQL using additional psychometric analysis methods. We will use methods such as item tree analysis to characterize students' knowledge states at various points within the physics curriculum and explore the potential hierarchy of skills measured by the PIQL \cite{young2018,Schrepp1999,Sargin2009}. We will also modify polytomous models of item response theory for analysis of responses to MCMR items \cite{Bock1972,Bock2007,Suh2010}. Such methods will extend our use of item response curves, and could lead to a more robust multi-level scoring scheme from which reasoning development patterns are likely to emerge.

%Mapping Growth of PQL paragraph
Lastly, we note that poorly developed PQL affects students in subsequent course-taking. Recent research suggests that students enrolled in advanced physics courses, which make use of mathematics such as linear algebra and differential equations, do not necessarily understand the underlying rationale for the mathematical procedures used to solve problems \cite{caballero2015,Smith2013,Smith2015b}. Many students learn procedures effectively. Instructors, as well as the students themselves, would like for these procedures to also be conceptualized \cite{caballero2015}. For many students, however, quantitative literacy is not a strong outcome of prior physics instruction. Our hope is that the PIQL will serve as a valuable tool for tracking and improving the development of physics quantitative literacy throughout the sequence of undergraduate physics instruction.

%THIS IS THE END OF THE PAPER

% Except for this part.
\begin{acknowledgements}
The authors thank Steve Kanim for valuable discussions of and contributions to foundational work on students' proportional and covariational reasoning, Michael Loverude for contributions to our investigations of covariational reasoning, and Brian Stephanik for insightful feedback. We also thank our project advisory board members: Jessica Cohen, Glen Davenport, Lin Ding, and Michael Oehrtman. We are indebted to the UW Physics Education Group for supporting this work in ways too varied to list here, and most especially to Lillian McDermott, posthumously, for helping build a field that takes student learning to heart. This work was partially supported by the National Science Foundation under grants DUE-1832836, DUE-1832880, DUE-1833050, and DGE-1762114.
\end{acknowledgements}

\bibliography{manualBib.bib}

% And a little bit more down here.
\end{document}